\documentclass[preprint]{aastex}

\shorttitle{IRAS17149-3916}
\shortauthors{Roman-Lopes et al.}

\begin{document}


\title{The young massive stellar cluster associated to RCW121}


\author{A. Roman-Lopes\altaffilmark{1} and Z. Abraham\altaffilmark{1}}





\altaffiltext{1}{Instituto de Astronomia, Geof\'\i sica e Ci\^encias Atmosf\'ericas, 
Universidade de
S\~ao Paulo \\ Rua do Mat\~ao 1226, 05508-900, S\~ao Paulo, SP, Brazil}


\begin{abstract}

We report NIR broad and narrow band photometric observations in the direction of the 
IRAS17149-3916 source that reveal the presence of a young cluster of  massive stars embedded in 
an HII region coincident with RCW121. These observations, together with published radio data, 
MSX and Spitzer images were used to determine some of the physical parameters of the region. We 
found 96 cluster member candidates in an area of about 1.5 $\times$ 2.0 square arcmin, 30\% of 
them showing excess emission in the NIR. IRS 1, the strongest source in the cluster with an 
estimated spectral type of O5V-O6V ZAMS based on the color-magnitude diagram, is probably the 
main ionizing source of the HII region detected at radio wavelengths.
Using the integrated Br$\gamma$  and the 5 GHz flux densities, we derived a mean visual 
extinction $A_V=5.49\pm ^{2.06}_{1.32}$ magnitudes. From the observed size of the Br$\gamma$ 
extended emission, we calculated the emission measure $E= 4.5\times 10^{24}$cm$^{-5}$ and the 
electron density $n_e=2.6\times 10^3$cm$^{-3}$, characteristic of compact HII regions. 

\end{abstract}





\section{Introduction}

Young massive stars are found inside dense molecular clouds generally suffering high extinction 
at optical wavelengths, mainly due to the large amount of gas and dust remaining from the star 
forming process \citep{churchwell90,cesaroni91}. Color-color and color-magnitude diagrams from 
near infrared photometry (NIR) allow the determination of the reddening of each individual star 
(at least for those that do not present intrinsec excess emission in the NIR). 
Longslit $K$-band spectra have shown that a large number of those NIR sources present broad 
Br$\gamma$ emission and no photospheric features \citep{armand96, hanson02}, typical  of very 
young pre-main sequence stars surrounded by disks and/or cocoons \citep{churchwell90}. 

On the other hand, narrow band imaging observations of star forming regions (e.g. centered in 
the Br$\gamma$ line and in the adjacent continuum) are powerful tools to study the emission from 
the extended surroundings. In fact, by comparing the Br$\gamma$ flux density with the free-free 
radio emission, it is possible to estimate the absorption \citep{hanson02}. Moreover, using the 
measured size of the  Br$\gamma$ emitting region and the low resolution radio flux density, it 
is possible to obtain the emission measure and mean value for the electron density 
\citep{mezger67}. We have to take into account that the measured size of the Br$\gamma$ emitting 
region is limited by the sensitivity of the NIR detector systems. The recent near to mid 
infrared images obtained by the Spitzer space telescope 
\footnote{http://irsa.ipac.caltech.edu/Missions/spitzer.html}, represent an additional constrain 
to the size of the ionized region, since they delineate the dusty border of the HII region.

RCW121 located at l=$348\fdg 2$ and b=-1$\fdg 0$ \citep{rod60} is a strong (11.9 Jy) and compact 
(about 1 arcmin angular diameter) 5 GHz  continuum source (PMNJ 1718-3916), observed  in the 
H109$\alpha$ and H110$\alpha$ recombination lines by \citet{caswell87}. 
The CS(2-1) transition, typical of dense molecular clouds, was also detected by 
\citet{bronfman96} and  methanol maser emission at 6.7 GHz, generally associated to high mass 
stellar forming regions, was observed with the Parkes radio telescope by \citet{walsh97}.
Maser emission from excited OH in the $^{2}\Pi_{1/2}$ J=1/2 state was found by \citet{cohen95} 
at all three transitions (4765, 4750 and 4660 MHz). 
\citet{dutra03}, by visual inspection of 2MASS images, found a cluster of stars in the direction 
of RCW121; this fact, together with the presence of an IRAS source with colors of compact HII 
regions \citep{wood89} indicates that star formation has  recently occurred. 
Despite of the overwhelming evidences pointing  to the existence of young and massive stellar 
objects associated with RCW121, no detailed studies of the stellar content of the region were 
made up to the present.

This work is part of a larger survey \citep{roman03,roman04a,roman04b}, aimed at the detection 
of the exciting stars of compact HII regions associated with IRAS point sources which are also 
strong CS emiters \citep{bronfman96}.
In the next sections we present the results of NIR broad and narrow band photometric 
observations of a young cluster of  massive stars found in the direction of the IRAS17149-3916 
source.
In Section 2 we describe the NIR observations; in Section 3 we present the results from the  
broad and narrow band NIR filters and in Section 4 we summarize the work. 

\section{Observations and data reduction}

The NIR imaging photometric observations were performed in  2003 May in the direction of IRAS 
point source 17149-3916,  using the near infrared camera of the Laborat\'orio Nacional de 
Astrofisica, Brazil, mounted on the 0.6 m Boller \& Chivens (BC) telescope; the resulting plate 
scale of the images is 0.48$"$pixel$^{-1}$.
The images were made using the $\it{J}$ and $\it{H}$ broad band filters and the $\it{K}$ narrow 
band filter $\it{C1}$,  centered at $2.14 \mu$m  continuum.
Since meaningful information about the ionized regions can be obtained  from  narrow band 
Br$\gamma$ images \citep{roman05}, we also made observation in
2005 May, using the near infrared camera, mounted on the 1.6 m Perkin-Elmer (PE) telescope; the 
resulting plate scale of the images is 0.24$"$pixel$^{-1}$.

The observations (mean FWHM values of the  point-spread functions, the total integration times, 
etc ) are summarized in Table 1. The NIR images were reduced using programs within the IRAF 
image reduction package system. Each image was corrected for nonlinear response, 
dark-subtracted, flattened, and masked to eliminate bad pixels. We observed calibration stars 
from the list of 2MASS standards at similar air mass of our science targets. In addition we 
checked our photometric results by comparing it with the 2MASS photometry for common bright 
field stars.  Details about the calibration and reduction procedures can be found in 
\citet{roman03}.

\begin{deluxetable}{cccccc}
\tabletypesize{\scriptsize}
\tablecaption{Summary of the NIR observations \label{tbl-1}}
\tablewidth{0pt}
\tablehead{
\colhead{Filter}&\colhead{Date}&\colhead{Telescope}&
\colhead{ITIME}&
\colhead{FWHM($"$)}&
}
\startdata

\hline

$J$& May 2003 & BC \footnote{}& 840 & 1.9 &\\
$H$& May 2003 & BC & 420 & 2.0 &\\
$K(C1)$& May 2003 & BC & 7200 & 2.1 &\\
$K$(Br$\gamma$)& May 2005 & PE \footnote{} & 360 & 1.7 &\\
$K$(C1)& May 2005 & PE  & 360 & 1.7 &\\
\hline

\enddata

{$^1$} {Boller $\&$ Chivens 0.6 m telescope}

{$^2$} {Perkin $\&$ Elmer 1.6 m telescope}



\end{deluxetable}

\section{Results}

Figure 1 shows the $J$, $H$ and $K$ near infrared  images obtained during the 2003 observing run 
(upper, central and lower panels, respectively).

\subsection{The stellar cluster}
To separate the cluster population from the field stars we analyzed the stellar colors in two 
distinct areas as can be seen in Figure 1: one, labeled "cluster" and measuring about 2.2' x 
2.4', encloses the IR nebula;  the other, labeled "control" and measuring about 8.8' x 1.3', 
includes the eastern and southern parts of the images. 
From the ($J-H$) versus ($H-K$) comparative diagrams, shown in Figure 2,  we can see that a 
significant number of "cluster" sources  have excess emission at the longer wavelengths, 
probably due to the presence of warm circumstellar dust or disks \citep{lada92}. On the other 
hand, the majority of the sources in the "control"  diagram lie inside  the reddening band for 
the Galactic field stars. They can be divided in two groups separated by a gap around $(J-H) 
\sim 1.4$: the first is formed by foreground field stars while the other is composed probably by 
giants stars located behind the molecular cloud.
  
We assumed as candidate cluster members all NIR sources that lie along the early type reddening 
band and those showing excess emission in the "cluster" color-color diagram. 
For the infrared sources not detected in the $J$ band a different criteria had to be used. In 
order to distinguish the genuine candidate cluster members from the field stars, we analyzed for 
the two selected areas, the number of stars  N$(H-K)$ with $(H-K)$ colors  as a function of 
$(H-K)$. The results with the counts normalized to the cluster region area are shown in Figure 
3; we can see  in the cluster area an excess of counts in the color ranges $(H-K)\sim 0-0.75$ 
and  $(H-K)\geq 1$ . The first range corresponds to the cluster  sources and the second group 
can be formed either by genuine cluster members or by giant stars located behind the molecular 
cloud. For this reason, only six stars not detected in the $J$ band, with colors $(H-K)\sim 
0-0.75$ or $(H-K)\geq 1$ and located inside the nebula region were taken as candidates cluster 
members.

In Figure 4, we show the $K$ band contour diagram in which  the relative position of each 
selected source is indicated. There, we show the objects that present excess emission in the NIR 
(filled triangles), those without (open triangles), as well as the  sources detected only in the 
$H$ and $K$ bands (filled squares).

We assumed a kinematic distance to the cluster of 1.6 kpc \citep{walsh97}, for a galactocentric 
distance of 8.5 kpc. 
Using this value, the luminosity of the stars in Table 2 can be inferred from the $J$ versus 
$(J-H)$ color-magnitude diagram  shown in Figure 5.
The position of the main-sequence is  plotted together with each spectral type earlier than F8V. 
The intrinsic colors were taken from \citet{koorneef83} and the absolute $J$ magnitudes were 
computed from the absolute visual luminosity for ZAMS given by \citet{hanson97}. 
The reddening vector for a ZAMS B0 V star, as in \citet{fitzpatrick99}, is shown by the dashed 
line together with the positions of visual extinction $A_V = 10$, 20 and 30 magnitudes indicated 
by filled circles. 

From the color-magnitude diagram, we can see that IRS1, IRS5, IRS9 and IRS12 are the most 
luminous objects without excess emission in the whole region, being the best ionizing source 
candidates. An inspection of the relative position of these sources shows that only IRS1 is 
located in the central part of the infrared nebula; the others lie at the border of the extended 
emission region.  In this sense IRS1,  an O5V-O6V ZAMS from Figure 5, probably is the main 
source of Lyman continuum photons that are ionizing the HII region detected at radio 
wavelengths.

\begin{deluxetable}{crrrrrr}
\tabletypesize{\scriptsize}
\tablecaption{List of the selected near-infrared sources \label{tbl-2}}
\tablewidth{0pt}
\tablehead{
\colhead{IRS}&\colhead{$\alpha$(J2000)}&\colhead{$\delta$(J2000)}&
\colhead{$J_{CamIV}$}&
\colhead{$H_{CamIV}$}&\colhead{$K_{CamIV}$}&
}
\startdata

\hline

1   &   17:18:25.51&	-39:19:08.8&	8.93 $\pm$0.02\phn&	8.36 $\pm$0.02\phn&	8.02 
$\pm$0.02\phn \\
2   &   17:18:25.73&	-39:18:27.0&	11.81 $\pm$0.02\phn& 9.64 $\pm$0.02\phn&	8.45 
$\pm$0.02\phn\\
3   &   17:18:22.27&	-39:18:42.5&	11.60 $\pm$0.02\phn& 10.23 $\pm$0.02\phn&	9.06 
$\pm$0.02\phn\\
4   &   17:18:22.92&	-39:18:22.7&	15.03 $\pm$0.03\phn& 11.70 $\pm$0.02\phn&	9.72 
$\pm$0.02\phn\\
5   &   17:18:20.59&	-39:18:07.2&	11.87 $\pm$0.03\phn& 10.66 $\pm$0.02\phn&	10.06 
$\pm$0.02\phn\\
6   &   17:18:30.72&	-39:18:40.0&	11.43 $\pm$0.02\phn& 10.75 $\pm$0.02\phn&	10.26 
$\pm$0.02\phn\\
7   &   17:18:25.66&	-39:19:09.8&	11.61 $\pm$0.03\phn& 10.86 $\pm$0.03\phn&	10.32 
$\pm$0.04\phn\\
8   &   17:18:26.62&	-39:17:51.7&	12.05 $\pm$0.02\phn& 10.98 $\pm$0.03\phn&	10.40 
$\pm$0.02\phn\\
9   &   17:18:25.34&	-39:18:24.8&	13.53 $\pm$0.03\phn& 11.47 $\pm$0.02\phn&	10.42 
$\pm$0.02\phn\\
10   &   17:18:25.32&	-39:19:09.5&	11.87 $\pm$0.03\phn& 11.35 $\pm$0.02\phn&	10.60 
$\pm$0.08\phn\\
11   &   17:18:28.75&	-39:19:15.6&	11.23 $\pm$0.02\phn& 10.96 $\pm$0.02\phn&	10.69 
$\pm$0.03\phn\\
12   &   17:18:30.36&	-39:19:53.0&	14.97 $\pm$0.04\phn& 12.16 $\pm$0.02\phn&	10.75 
$\pm$0.03\phn\\
13   &   17:18:30.70&	-39:19:52.7&	11.52 $\pm$0.03\phn& 11.12 $\pm$0.02\phn&	10.86 
$\pm$0.03\phn\\
14   &   17:18:25.39&	-39:19:12.0&	12.30 $\pm$0.03\phn& 11.36 $\pm$0.03\phn&	10.91 
$\pm$0.03\phn\\
15   &   17:18:31.90&	-39:19:40.4&	12.27 $\pm$0.03\phn& 11.33 $\pm$0.03\phn&	10.93 
$\pm$0.03\phn\\
16   &   17:18:32.04&	-39:17:54.2&	13.87 $\pm$0.03\phn& 12.06 $\pm$0.03\phn&	11.08 
$\pm$0.03\phn\\
17   &   17:18:31.68&	-39:18:24.1&	12.28 $\pm$0.03\phn& 11.48 $\pm$0.03\phn&	11.13 
$\pm$0.03\phn\\
18   &   17:18:25.18&	-39:18:46.8&	14.87 $\pm$0.03\phn& 12.44 $\pm$0.03\phn&	11.20 
$\pm$0.03\phn\\
19   &   17:18:24.48&	-39:19:39.7&	11.70 $\pm$0.03\phn& 11.43 $\pm$0.03\phn&	11.22 
$\pm$0.03\phn\\
20   &   17:18:22.42&	-39:18:21.2&	12.77 $\pm$0.03\phn& 11.74 $\pm$0.03\phn&	11.21 
$\pm$0.03\phn\\
21   &   17:18:32.98&	-39:19:32.2&	12.04 $\pm$0.03\phn& 11.58 $\pm$0.03\phn&	11.27 
$\pm$0.03\phn\\
22   &   17:18:25.75&	-39:19:55.9&	14.35 $\pm$0.03\phn& 12.33 $\pm$0.03\phn&	11.41 
$\pm$0.03\phn\\
23   &   17:18:21.70&	-39:19:15.2&	14.78 $\pm$0.03\phn& 12.66 $\pm$0.03\phn&	11.45 
$\pm$0.03\phn\\
24   &   17:18:21.60&	-39:18:31.7&	14.91 $\pm$0.03\phn& 12.63 $\pm$0.03\phn&	11.46 
$\pm$0.03\phn\\
25   &   17:18:29.40&	-39:17:56.0&	12.30 $\pm$0.03\phn& 11.78 $\pm$0.03\phn&	11.47 
$\pm$0.03\phn\\
26   &   17:18:30.91&	-39:18:33.8&	12.66 $\pm$0.03\phn& 11.89 $\pm$0.03\phn&	11.49 
$\pm$0.03\phn\\
27   &   17:18:20.59&	-39:18:13.0&	12.02 $\pm$0.03\phn& 11.72 $\pm$0.03\phn&	11.50 
$\pm$0.03\phn\\
28   &   17:18:22.44&	-39:19:20.6&	12.89 $\pm$0.03\phn& 12.07 $\pm$0.03\phn&	11.52 
$\pm$0.03\phn\\
29   &   17:18:25.97&	-39:19:26.8&	13.48 $\pm$0.03\phn& 12.29 $\pm$0.03\phn&	11.59 
$\pm$0.03\phn\\
30   &   17:18:26.16&	-39:17:42.7&	15.25 $\pm$0.03\phn& 12.88 $\pm$0.03\phn&	11.61 
$\pm$0.04\phn\\
31   &   17:18:23.33&	-39:19:08.0&	13.45 $\pm$0.03\phn& 12.35 $\pm$0.03\phn&	11.76 
$\pm$0.03\phn\\
32   &   17:18:31.37&	-39:18:10.4&	14.98 $\pm$0.03\phn& 12.84 $\pm$0.03\phn&	11.84 
$\pm$0.03\phn\\
33   &   17:18:23.83&	-39:19:04.4&	12.98 $\pm$0.03\phn& 12.31 $\pm$0.03\phn&	11.86 
$\pm$0.03\phn\\
34   &   17:18:29.69&	-39:19:02.3&	12.74 $\pm$0.03\phn& 12.23 $\pm$0.03\phn&	11.90 
$\pm$0.03\phn\\
35   &   17:18:25.54&	-39:19:23.2&	13.55 $\pm$0.03\phn& 12.48 $\pm$0.03\phn&	11.90 
$\pm$0.03\phn\\
36   &   17:18:29.81&	-39:19:07.3&	15.62 $\pm$0.04\phn& 13.21 $\pm$0.03\phn&	11.91 
$\pm$0.03\phn\\
37   &   17:18:21.48&	-39:19:13.8&	13.05 $\pm$0.03\phn& 12.39 $\pm$0.03\phn&	11.92 
$\pm$0.03\phn\\
38   &   17:18:27.89&	-39:17:58.6&	15.28 $\pm$0.04\phn& 13.02 $\pm$0.03\phn&	11.93 
$\pm$0.03\phn\\
39   &   17:18:21.77&	-39:19:22.1&	15.36 $\pm$0.05\phn& 13.24 $\pm$0.03\phn&	11.93 
$\pm$0.03\phn\\
40   &   17:18:24.48&	-39:19:21.4&	13.28 $\pm$0.03\phn& 12.51 $\pm$0.03\phn&	11.96 
$\pm$0.03\phn\\
41   &   17:18:21.22&	-39:18:38.5&	14.86 $\pm$0.04\phn& 13.13 $\pm$0.04\phn&	12.01 
$\pm$0.05\phn\\
42   &   17:18:25.87&	-39:18:56.9&	13.82 $\pm$0.04\phn& 12.71 $\pm$0.03\phn&	12.09 
$\pm$0.05\phn\\
43   &   17:18:24.70&	-39:18:41.8&	12.71 $\pm$0.04\phn& 12.25 $\pm$0.03\phn&	12.03 
$\pm$0.05\phn\\
44   &   17:18:25.30&	-39:18:37.8&	13.36 $\pm$0.04\phn& 12.62 $\pm$0.04\phn&	12.10 
$\pm$0.05\phn\\
45   &   17:18:25.75&	-39:19:25.3&	13.77 $\pm$0.04\phn& 12.81 $\pm$0.04\phn&	12.14 
$\pm$0.05\phn\\
46   &   17:18:22.90&	-39:19:40.8&	12.91 $\pm$0.04\phn& 12.46 $\pm$0.04\phn&	12.18 
$\pm$0.05\phn\\
47   &   17:18:23.33&	-39:18:27.4&	13.09 $\pm$0.04\phn& 12.61 $\pm$0.04\phn&	12.18 
$\pm$0.05\phn\\
48   &   17:18:23.35&	-39:18:56.2&	13.32 $\pm$0.04\phn& 12.68 $\pm$0.04\phn&	12.34 
$\pm$0.05\phn\\
49   &   17:18:24.48&	-39:18:09.7&	16.31 $\pm$0.06\phn& 13.82 $\pm$0.04\phn&	12.51 
$\pm$0.06\phn\\
50   &   17:18:26.57&	-39:19:03.4&	13.13 $\pm$0.04\phn& 12.79 $\pm$0.04\phn&	12.55 
$\pm$0.06\phn\\
51   &   17:18:24.14&	-39:18:31.0&	13.40 $\pm$0.04\phn& 12.88 $\pm$0.04\phn&	12.62 
$\pm$0.06\phn\\
52   &   17:18:32.74&	-39:18:14.0&	12.83 $\pm$0.04\phn& 12.45 $\pm$0.04\phn&	12.33 
$\pm$0.05\phn\\
53   &   17:18:23.28&	-39:19:05.9&	15.11 $\pm$0.05\phn& 13.74 $\pm$0.04\phn&	12.63 
$\pm$0.06\phn\\
54   &   17:18:21.94&	-39:18:41.4&	14.64 $\pm$0.04\phn& 13.37 $\pm$0.04\phn&	12.66 
$\pm$0.06\phn\\
55   &   17:18:32.88&	-39:19:05.5&	14.56 $\pm$0.04\phn& 13.37 $\pm$0.04\phn&	12.70 
$\pm$0.06\phn\\
56   &   17:18:28.32&	-39:19:49.8&	14.73 $\pm$0.04\phn& 13.46 $\pm$0.04\phn&	12.71 
$\pm$0.06\phn\\
57   &   17:18:23.47&	-39:19:52.3&	13.27 $\pm$0.04\phn& 12.93 $\pm$0.04\phn&	12.73 
$\pm$0.06\phn\\
58   &   17:18:20.62&	-39:19:55.2&	13.59 $\pm$0.04\phn& 13.06 $\pm$0.04\phn&	12.80 
$\pm$0.06\phn\\
59   &   17:18:24.74&	-39:19:06.6&	15.00 $\pm$0.05\phn& 13.63 $\pm$0.04\phn&	12.84 
$\pm$0.07\phn\\
60   &   17:18:25.61&	-39:19:16.7&	14.58 $\pm$0.04\phn& 13.31 $\pm$0.04\phn&	12.65 
$\pm$0.06\phn\\
61   &   17:18:24.43&	-39:18:40.7&	15.75 $\pm$0.05\phn& 14.00 $\pm$0.04\phn&	12.86 
$\pm$0.07\phn\\
62   &   17:18:24.43&	-39:19:23.5&	15.87 $\pm$0.07\phn& 14.70 $\pm$0.04\phn&	12.94 
$\pm$0.08\phn\\
63   &   17:18:24.48&	-39:19:10.9&	15.34 $\pm$0.06\phn& 13.80 $\pm$0.04\phn&	13.01 
$\pm$0.08\phn\\
64   &   17:18:21.29&	-39:18:51.1&	14.42 $\pm$0.04\phn& 13.40 $\pm$0.04\phn&	12.98 
$\pm$0.08\phn\\
65   &   17:18:30.58&	-39:18:37.1&	14.01 $\pm$0.04\phn& 13.49 $\pm$0.04\phn&	13.06 
$\pm$0.08\phn\\
66   &   17:18:22.90&	-39:18:59.0&	15.18 $\pm$0.04\phn& 13.81 $\pm$0.04\phn&	13.11 
$\pm$0.08\phn\\
67   &   17:18:23.52&	-39:19:10.6&	14.10 $\pm$0.04\phn& 13.51 $\pm$0.04\phn&	13.12 
$\pm$0.08\phn\\
68   &   17:18:26.35&	-39:19:18.5&	14.73 $\pm$0.04\phn& 13.64 $\pm$0.04\phn&	13.15 
$\pm$0.08\phn\\
69   &   17:18:21.05&	-39:18:44.6&	17.12 $\pm$0.14\phn& 14.53 $\pm$0.05\phn&	13.20 
$\pm$0.09\phn\\
70   &   17:18:21.77&	-39:18:48.2&	13.80 $\pm$0.04\phn& 13.42 $\pm$0.04\phn&	13.26 
$\pm$0.09\phn\\
71   &   17:18:26.86&	-39:18:49.0&	14.06 $\pm$0.04\phn& 13.61 $\pm$0.04\phn&	13.27 
$\pm$0.09\phn\\
72   &   17:18:33.19&	-39:18:13.7&	13.59 $\pm$0.04\phn& 13.27 $\pm$0.04\phn&	13.11 
$\pm$0.08\phn\\
73   &   17:18:23.28&	-39:18:46.1&	15.43 $\pm$0.06\phn& 14.29 $\pm$0.04\phn&	13.27 
$\pm$0.09\phn\\
74   &   17:18:29.86&	-39:19:28.9&	14.48 $\pm$0.04\phn& 13.76 $\pm$0.04\phn&	13.31 
$\pm$0.09\phn\\
75   &   17:18:27.14&	-39:19:11.3&	14.90 $\pm$0.05\phn& 13.96 $\pm$0.04\phn&	13.39 
$\pm$0.09\phn\\
76   &   17:18:22.75&	-39:18:45.4&	15.85 $\pm$0.06\phn& 14.49 $\pm$0.05\phn&	13.32 
$\pm$0.09\phn\\
77   &   17:18:19.94&	-39:17:57.1&	15.22 $\pm$0.05\phn& 14.26 $\pm$0.04\phn&	13.76 
$\pm$0.10\phn\\
78   &   17:18:32.38&	-39:19:10.9&	17.01 $\pm$0.08\phn& 14.97 $\pm$0.06\phn&	13.34 
$\pm$0.09\phn\\
79   &   17:18:22.75&	-39:18:11.5&	14.61 $\pm$0.05\phn& 13.92 $\pm$0.04\phn&	13.55 
$\pm$0.09\phn\\
80   &   17:18:26.50&	-39:19:32.9&	14.54 $\pm$0.05\phn& 13.95 $\pm$0.04\phn&	13.65 
$\pm$0.10\phn\\
81   &   17:18:22.18&	-39:19:26.0&	14.97 $\pm$0.05\phn& 14.22 $\pm$0.04\phn&	13.59 
$\pm$0.10\phn\\
82   &   17:18:26.40&	-39:19:06.6&	17.02 $\pm$0.13\phn& 15.67 $\pm$0.05\phn&	14.27 
$\pm$0.16\phn\\
83   &   17:18:23.33&	-39:18:39.2&	14.95 $\pm$0.05\phn& 14.42 $\pm$0.05\phn&	13.71 
$\pm$0.11\phn\\
84   &   17:18:30.74&	-39:19:14.5&	15.45 $\pm$0.05\phn& 14.65 $\pm$0.04\phn&	13.75 
$\pm$0.10\phn\\
85   &   17:18:31.15&	-39:19:52.0&	13.84 $\pm$0.04\phn& 13.50 $\pm$0.04\phn&	13.32 
$\pm$0.09\phn\\
86   &   17:18:21.55&	-39:19:48.0&	15.31 $\pm$0.05\phn& 14.42 $\pm$0.05\phn&	13.82 
$\pm$0.11\phn\\
87   &   17:18:28.94&	-39:18:44.6&	15.05 $\pm$0.04\phn& 14.32 $\pm$0.04\phn&	13.83 
$\pm$0.12\phn\\
88   &   17:18:26.30&	-39:19:50.2&	14.67 $\pm$0.04\phn& 14.24 $\pm$0.05\phn&	14.01 
$\pm$0.13\phn\\
89   &   17:18:22.42&	-39:18:36.4&	15.44 $\pm$0.04\phn& 14.06 $\pm$0.04\phn&	13.28 
$\pm$0.09\phn\\
90   &   17:18:28.80&	-39:19:54.1&	14.33 $\pm$0.04\phn& 13.95 $\pm$0.04\phn&	13.88 
$\pm$0.11\phn\\
91   &   17:18:30.48&	-39:18:45.7&	\phn& 14.11 $\pm$0.05\phn&	11.31 $\pm$0.04\phn\\
92   &   17:18:28.94&	-39:19:21.7&	\phn& 13.25 $\pm$0.03\phn&	11.34 $\pm$0.04\phn\\
93   &   17:18:22.1&	-39:19:11.6&	\phn& 14.68 $\pm$0.05\phn&	12.89 $\pm$0.07\phn\\
94   &   17:18:21.36&	-39:19:54.1&	\phn& 15.60 $\pm$0.06\phn&	13.81 $\pm$0.11\phn\\
95   &   17:18:30.53&	-39:19:37.6&	\phn& 14.33 $\pm$0.05\phn&	13.82 $\pm$0.11\phn\\
96   &   17:18:26.59&	-39:17:56.8&	\phn& 13.98 $\pm$0.05\phn&	13.37 $\pm$0.09\phn\\

\hline

\enddata



\end{deluxetable}

\subsection{The IRAS source} \label{bozomath}

In Figure 6 we show the  $K$ band image ($C1$ filter) of the region around IRAS 17149-3916, 
overlapped  by  the IRAS error ellipse (broken line). 
Several stars fall inside the  ellipse; a more accurate coordinate for IRAS 17149-3916 was 
obtained from the Midcourse Space Experiment \citep{price01} - 
MSX\footnote{http://www.ipac.caltech.edu/ipac/msx/msx.html} point source catalog.  
We found one MSX source within the IRAS error ellipse (G348.2362-0.9809), with coordinates 
$\alpha(\rm{J2000)=17^{h}18^{m}25.2^{s}}$, $\delta(\rm{J2000)=-39^{d}19^{m}42^{s}}$, which is 
shown as a contour diagram in Figure 6. None of the stars coincides with the MSX coordinates. In 
fact, from figure 6, we can see that probably several individual sources can be contributing to 
the IRAS emission. 
 
 \subsection{ Br$\gamma$ emission from the infrared nebula}

As we already mentioned, there is a strong and compact (with about 1 arcmin radius) radio source 
in the direction of IRAS17149-3916 (PMNJ 1718-3918), with a flux density  of 11.9 Jy  at 
4.85GHz. In order to study this extended emission, we re-observed in the 2005 observing run, the 
field toward IRAS17149-3916 source trough the narrow band $K$ filters, which are centered in the 
Br$\gamma$ line and in the adjacent continuum (see table 1).

To calculate the total Br$\gamma$ emission, we first scaled  the counts of the continuum image 
($C1$ filter) to the same level of the Br$\gamma$ image, using the common bright field stars.
In Figure 7, we show the Br$\gamma$ contour diagram constructed from the difference between the 
Br$\gamma$ and $C1$ images, overlaying the 
Spitzer\footnote{http://www.ipac.caltech.edu/data/SPITZER/SPITZER} 3.6$\mu$m image of the region 
around IRAS17149-3916; we also indicated the position of four prominent Spitzer sources 
coincident with the infrared objects IRS1, IRS2, IRS3 and IRS4 from our survey. From that 
figure, we can also see that the Br$\gamma$ emission presents a "shell" like shape, which seems 
to be enclosed  by 3.6 $\mu$m dust emission. From the spectral type of IRS1 (O5V-O6V derived 
from the color-magnitude diagram in figure 5), which is the most luminous source in the whole 
region, we can speculate that the observed Br$\gamma$ flux density probably is generated by the 
same ionized gas that produces the free-free emission detected at radio wavelenghts.

Next we constructed the flux calibrated contour diagrams of the two images (the Br$\gamma$ and 
$C1$ contour diagrams) and, by measuring the area between contours, we obtained the net 
Br$\gamma$ flux density using the relation between magnitude and flux density given by Koorneef 
(1983), finding  an integrated Br$\gamma$ flux density of $7.1(\pm 1.1$)$\times$10$^{-11}$ erg 
cm$^{-2}$ s$^{-1}$.

\subsection{ Mean visual extinction and  emission measure determined from the Br$\gamma$ and 
radio flux densities}

To estimate the mean visual extinction $A_V$ in the direction of the nebula, we first used the 
integrated Br$\gamma$  flux density, uncorrected for reddening, to compute the corresponding 
number of Lyman continuum photons, using the expression derived by \citet{ho90}:

\begin{equation}
N_{Ly}= 2.9\times 10^{45} \biggl(\frac{D}{\rm kpc}\biggr)^2 \biggl(\frac {3S_{\rm Br\gamma}}{\rm 
10^{-12}\,erg\, cm^{-2}\, s^{-1}}\biggr)\; \rm{s^{-1}} 
\end{equation}

We obtained (1.58$\pm 0.25$) $\times 10^{48}$ photons s$^{-1}$. We compared this result with 
that derived from the 6 cm flux density, using the relation from \citet{armand96}, assuming an 
electron temperature $T_e = 7500K$:

\begin{equation}
N_{Ly}({6\; \rm cm})= 5.25\times 10^{48} \biggl(\frac{D}{\rm kpc}\biggr)^2 T^{-0.45}S_{\rm 
5GHz}\;(\rm Jy)\; \rm{s^{-1}} 
\end{equation}

\clearpage

We found $N_{Ly}({\rm 6\; cm})=5.25\times 10^{48}$ photons s$^{-1}$, which corresponds to a mean 
visual extinction $A_V=5.49\pm ^{2.06}_{1.32}$ magnitudes.
This result agrees with that obtained from the color-magnitude diagram for the IRS1 source (see 
fig.4), which appears reddened by about $A_V \simeq 6.5$ mag.

We calculate the emission measure $E$ from the detected 5 GHz flux density, using the observed 
size of the Br$\gamma$ extended emission (about 55 arcsec) and the expression of the expected 
free-free emission from an optically thin plasma at wavelength $\lambda$, given by:

\begin{equation}
S_\nu = 5.4\times 10^{-16}g_{ff}(\lambda,T)\Omega E T^{-1/2}e^{-hc/\lambda kT} \; {\rm Jy}
\end{equation}

\noindent
where $E$ is the emission measure (cm$^{-5}$), $\Omega$ the solid angle of the source and 
$g_{ff}(\lambda,T)$ the Gaunt factor, which for radio wavelengths can be calculated from:

\begin{equation}
g_{ff}(\lambda, T) = \frac{\sqrt{3}}{\pi}\biggl[17.7+\ln
 \biggl(\frac{T^{3/2}\lambda}{c}\biggr)\biggr].
\end{equation}

Assuming $T$=7500K and $S_\nu$=11.9 Jy we found $E= 4.5\times 10^{24}$cm$^{-5}$, which results 
in a mean electron density $n_e=2.6\times 10^3$cm$^{-3}$, characteristic of compact HII regions 
\citep{Churchwell02}.

\subsection{ The IRS4 near infrared source}

In figure 7, we indicated the counterparts of the IRS1, IRS2, IRS3 and IRS4 sources in the $3.6 
\mu$m Spitzer image. As we have already mentioned,  IRS1  is the best candidate to be the main 
ionzing source in the cluster. The other three NIR sources present large excess of emission in 
the  color-color diagram, appearing also as very bright IR sources in the $3.6 \mu$m image.

We  attributed the excess emission from IRS2 and IRS3 to dust that probably surrounds these 
young stars. In the case of  IRS4, we found a residual in the difference between the point 
spread function photometry (Br$\gamma$- $C1$), showing the presence of Br$\gamma$ line emission, 
suggesting the presence of a highly embedded UCHII region. This can be seen in Figure 8, where 
we present the $K$(Br$\gamma$)-$K$(C1) versus $K$ magnitude diagram, constructed from the point 
spread function photometry of the 41 common bright stars (K $\leq$11.5) in the Br$\gamma$ and 
$C1$ narrow band images.  From that figure we can see that all stars but one (IRS4) have 
compatible $K$ photometry, considering the photometric errors.

\section{Conclusions}

NIR observations in the direction of RCW121 reveals the presence of a young cluster of  massive 
stars embedded in an HII region coincident with IRAS17149-3916 source. These observations, 
together with published radio data, MSX and Spitzer images were used to determine some of the 
physical parameters of the region. We found 96 cluster member candidates in an area of about 1.5 
$\times$ 2.0 square arcmin, 30\% of them showing excess emission in the NIR. IRS 1, the 
strongest source in the cluster with an estimated spectral type of O5V-O6V ZAMS based on the 
color-magnitude diagram, is probably the main ionizing source of the HII region detected at 
radio wavelengths.

We did not find a star directly associated with the MSX source coordinates; in fact, probably 
more than one  NIR source could be heating the dust, since the strongest of them are 
concentrated near the maximum of  the 8.28 $\mu$m emission in the MSX image. This result is 
different from that obtained by \citet{roman03,roman04a,roman04b}, in which a single and strong 
NIR source have coordinates that agree with the MSX source.

From narrow band Br$\gamma$ and continuum images we measured an extended Br$\gamma$  flux 
density of $(7.1 \pm 1.1$)$\times$10$^{-11}$ erg cm$^{-2}$ s$^{-1}$.
We found that the Br$\gamma$ emission presents a shell-like structure, which seems to be 
enclosed  by 3.6 $\mu$m dust emission.
Using the integrated Br$\gamma$  flux density, uncorrected for reddening, we computed the 
corresponding number of Lyman continuum photons and compared it with that obtained from the 5 
GHz flux density to derive a mean visual extinction of $A_V=5.49\pm ^{2.06}_{1.32}$ magnitudes. 
This result agrees with that inferred from the color-magnitude diagram for the IRS1 source, 
which appears reddened by about $A_V \simeq 6.5$ mag.
From the observed size of the Br$\gamma$ extended emission and the detected 5 GHz flux density, 
we calculated emission measure $E= 4.5\times 10^{24}$cm$^{-5}$ and  electron density 
$n_e=2.6\times 10^3$cm$^{-3}$, characteristic of compact HII regions \citet{Churchwell02}.
Moreover, the IRS4 object presents a residual in the difference between the point spread 
function photometry (Br$\gamma$- $C1$), showing the presence of  Br$\gamma$ line emission, 
suggesting the existence of a highly embedded UCHII region.

\section*{Acknowledgments}

This work was partially supported by the Brazilian agencies FAPESP and CNPq.
We acknowledge the staff of Laborat\'orio Nacional de 
Astrof\' \i sica for their efficient support.
This publication makes use of data products from the Two Micron All 
Sky Survey, which is a joint project of the University of 
Massachusets and the Infrared Processing and Analysis Center/California 
Institute of Technology, funded by the
National Aeronautics and Space Administration and the National Science Foundation. 
This research made use of data products from the Midcourse Space 
Experiment. This work is based [in part] on observations made with the Spitzer Space Telescope, 
which is operated by the Jet Propulsion Laboratory, California Institute of Technology under a 
contract with NASA.

  \begin{figure}
   \centering
   \includegraphics[bb=229 134 384 656,width=6.5cm,clip]{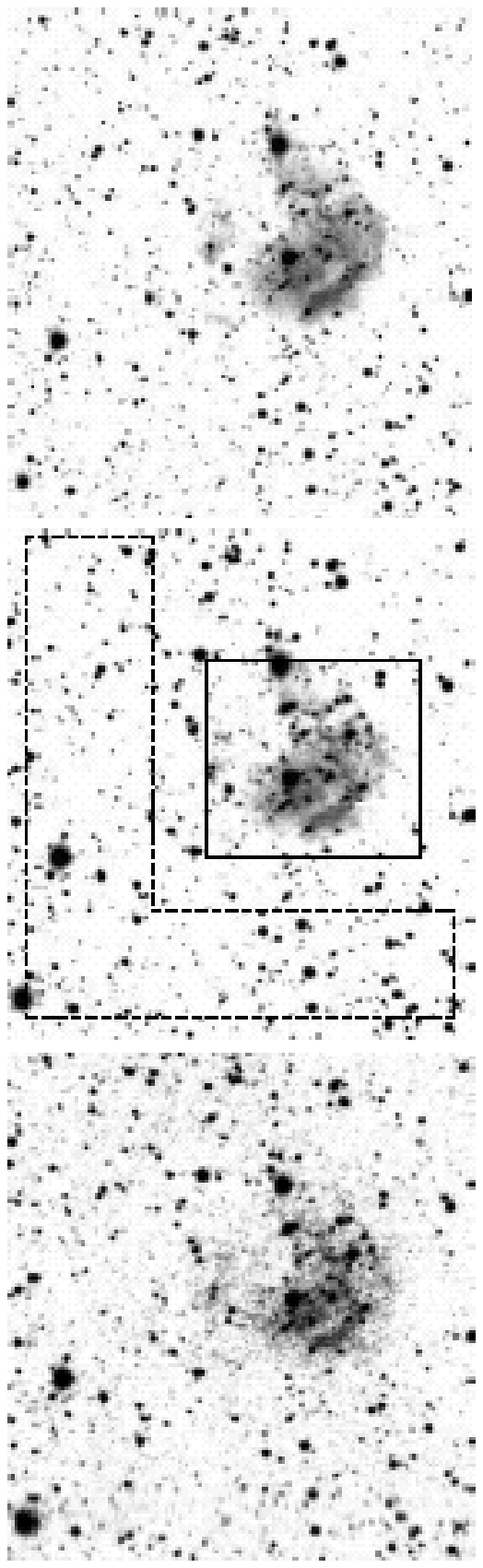}
      \caption{The $J$ (up), $H$ (center) and $K(C1)$ (down) infrared images (about 5.2' x 
5.6'). These frames were taken using the near infrared camera coupled to the 0.6 m Boller \& 
Chivens telescope, during the 2003 observing run. North is to the top, east to the left. The 
"cluster" and "control" regions are delimited by the dotted and dashed lines respectively.}
         \label{Fig1}
   \end{figure}
   
     \begin{figure}
   \centering
   \includegraphics[bb=191 112 412 671,width=7.5cm,clip]{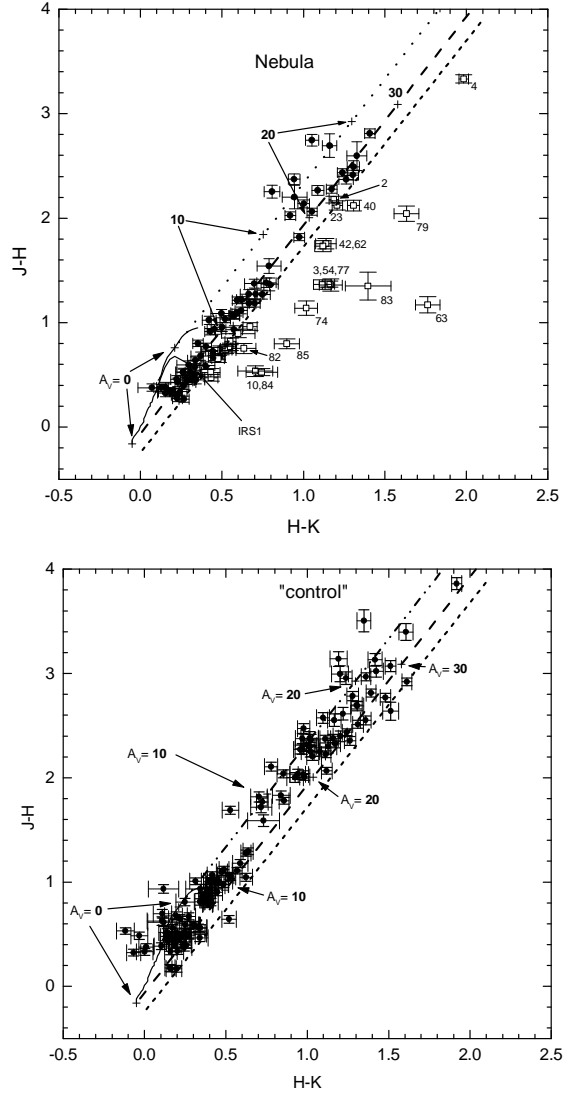}
      \caption{Color-color diagrams for two  comparative regions in our survey. The control 
diagram
probably contains only field objects; the nebula diagram has also some field stars 
but presents many young objects with excess emission in the NIR.
The locus of the main sequence and the giant branch are represented by the solid lines  
(Koornneef,1983), while the reddening vector for late and early type stars are represented by 
dotted and long dashed lines respectively. A third vector represented by the long dotted line 
indicate the place in the color-color diagram where the stars present excess emission in the 
NIR.}
         \label{Fig2}
   \end{figure}

\clearpage

  \begin{figure}
   \centering
   \includegraphics[bb=183 233 410 516,width=6cm,clip]{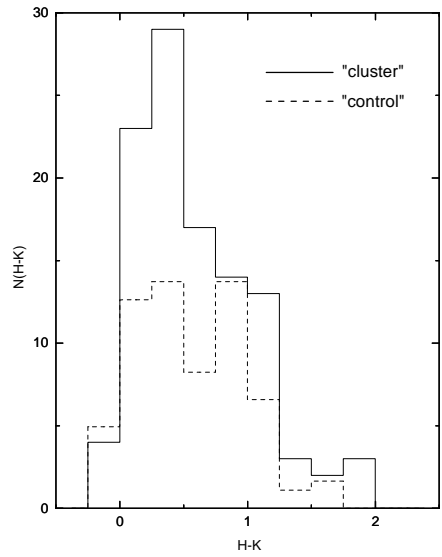}
      \caption{Comparative histogram of the distribution of the $\it{H-K}$ colors in the
control (dashed line) and in the cluster (solid line) regions. The counts of the control area 
were normalized to the area of the cluster region.}
         \label{Fig3}
   \end{figure}
   
   \clearpage

   \begin{figure}
   \centering
   \plotone{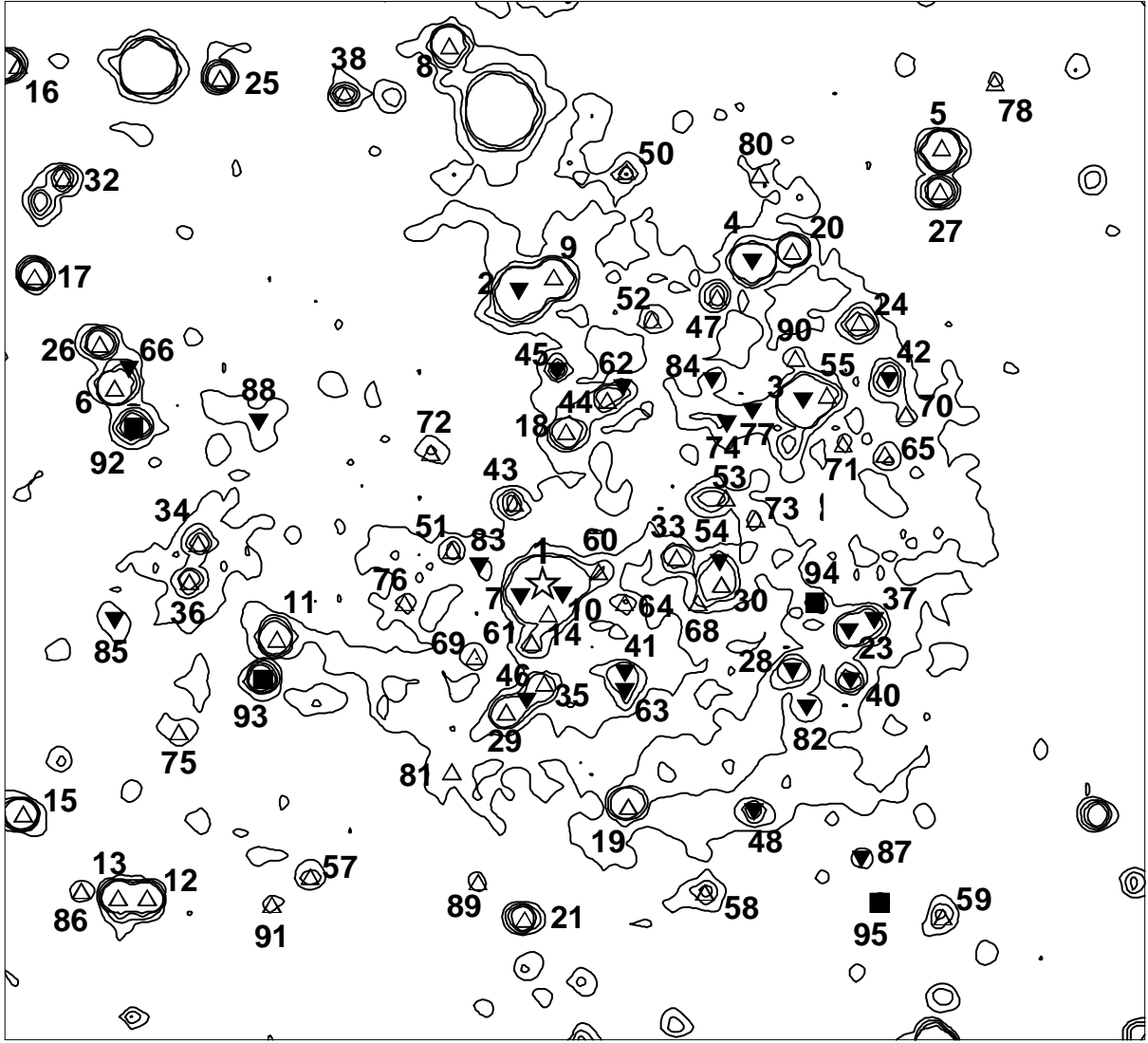}
      \caption{nb$K$-band contour diagram of the infrared nebula region.
The contours are spaced in intervals of 10\%, 40\%, 80\% and 100\% of the peak value.
Here we indicate the position of the sources with excess emission (filled triangles), those 
without (open triangles), the main ionizing source candidate (star), and  those only detected at 
$H$ and $K$ bands (filled squares).}
         \label{Fig4}
   \end{figure}

   \begin{figure}
   \centering
\plotone{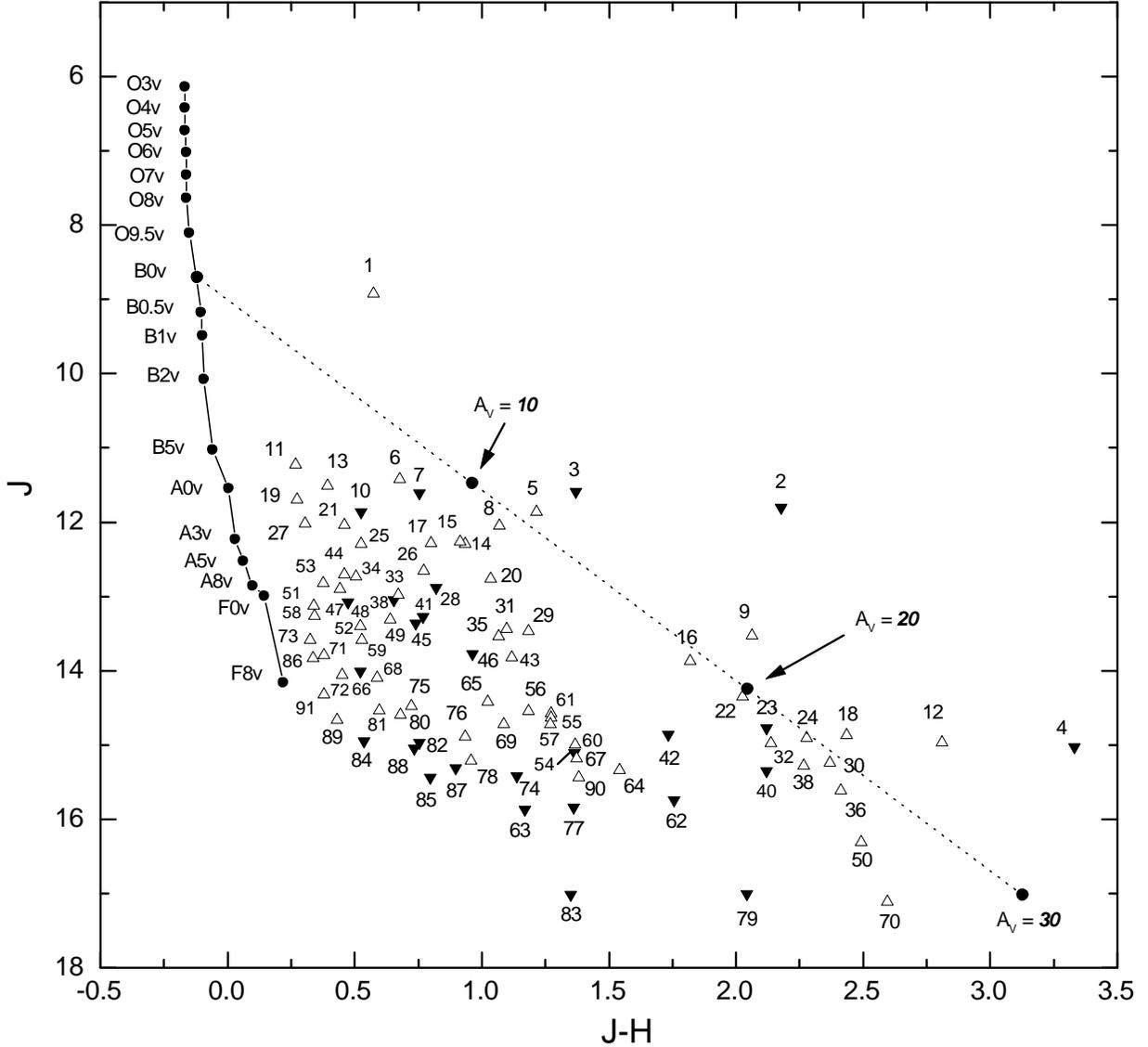}    
  \caption{The $J$ vs $J-H$ color-magnitude diagram of the sources in Table 1.
The locus of the main sequence at 1.6 kpc is shown by the solid line. The intrinsic colors are
from Koornneef (1983), while the absolute $J$ magnitudes were calculated from Hanson et al. 
(1997). The redening vector for a B0 ZAMS star ($dotted$ $line$) is
from Fitspatrick (1999). We also indicate the location ($bold$ $numbers$)
of $A_V$ = 10, 20 and 30 
mag of visual extinction as well as the sources that show excess (filled triangles) 
and those that do not (open triangles) in the color-color diagram.}
         \label{Fig5}
   \end{figure}
   
  \begin{figure}
   \centering
   \includegraphics[bb=165 252 426 545,width=8.5cm,clip]{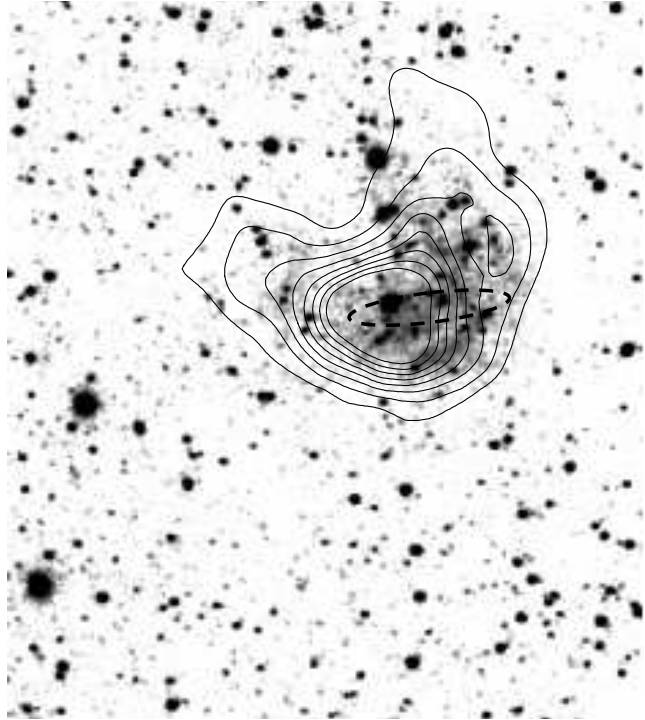}
      \caption{Contour diagram from the MSX A-band image (8.28 $\mu$m), overlayed by the LNA
$C1$ image. The IRAS coordinate error ellipse is represented by the dashed elipse. }
         \label{Fig6}
   \end{figure}
   
  \begin{figure}
   \centering
   \includegraphics[bb=129 225 482 522,width=8cm,clip]{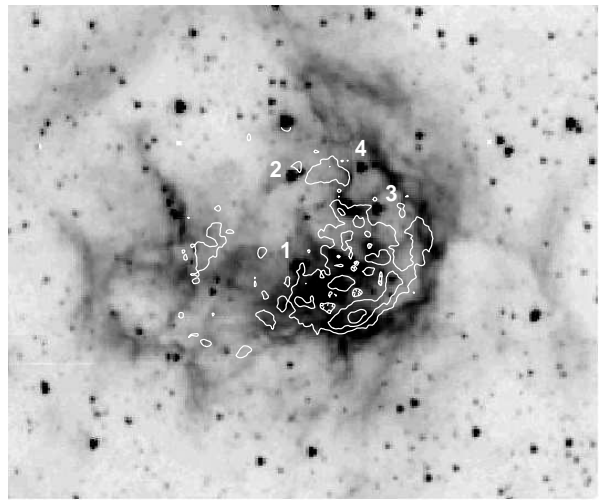}
      \caption{Contour Br$\gamma$ diagram overlayed by the Spitzer 3.6$\mu$m image of the region 
around IRAS17149-3916. The contours start at 695 counts arcsec$^{-1}$ and are spaced by the same 
value.}
         \label{Fig7}
   \end{figure}
   
\clearpage

  \begin{figure}
   \centering
   \includegraphics[bb=170 331 429 447,width=8cm,clip]{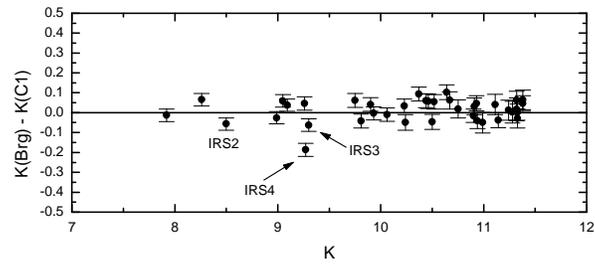}
      \caption{The $K$(Br$\gamma$)-$K$(C1) versus $K$ magnitude diagram, constructed from the 
PSF photometry of 41 common bright stars (K $\leq$11.5). The anomalous source in this diagram 
corresponds to the IRS4 source (see text).}
         \label{Fig8}
   \end{figure} 

\end{document}